\documentclass[a4paper]{article}

\usepackage{INTERSPEECH2020}
\usepackage{graphicx}

\title{WG-WaveNet: Real-Time High-Fidelity Speech Synthesis without GPU}
\name{Po-chun Hsu$^{1 2}$\thanks{This work was supported by NVIDIA and Taiwan AI Labs.}, Hung-yi Lee$^{1 2}$}
\address{$^1$College of Electrical Engineering and Computer Science, National Taiwan University\\
        $^2$Graduate Institute of Communication Engineering, National Taiwan University}
\email{f07942095@ntu.edu.tw, hungyilee@ntu.edu.tw}

\begin{document}

\maketitle
\begin{abstract}
  In this paper, we propose WG-WaveNet, a fast, lightweight, and high-quality waveform generation model.
  WG-WaveNet is composed of a compact flow-based model and a post-filter. The two components are jointly trained by maximizing the likelihood of the training data and optimizing loss functions on the frequency domains. 
  As we design a flow-based model that is heavily compressed, the proposed model requires much less computational resources compared to other waveform generation models during both training and inference time; even though the model is highly compressed, the post-filter maintains the quality of generated waveform.
  Our PyTorch implementation can be trained using less than 8 GB GPU memory and generates audio samples at a rate of more than 960 kHz on an NVIDIA 1080Ti GPU. Furthermore, even if synthesizing on a CPU, we show that the proposed method is capable of generating 44.1 kHz speech waveform 1.2 times faster than real-time. Experiments also show that the quality of generated audio is comparable to those of other methods.
  Audio samples are publicly available online.
\end{abstract}
\section{Introduction}

  Recently, neural network-based models have achieved state-of-the-art performance in speech tasks such as text-to-speech and voice conversion~\cite{shen2018natural, li2019neural, chou2018multi, qian2019autovc}. These models are typically composed of two parts. The first model conducts the speech tasks and generates a spectrogram~\cite{shen2018natural, qian2019autovc}, F0 frequencies, or other acoustic features~\cite{oord2016wavenet}. The second part, referred to as a vocoder, is a generative model or a heuristic method transforming acoustic features into audio samples. 

    WaveNet~\cite{oord2016wavenet} is first used as a neural vocoder to produce close-to-human natural speech~\cite{shen2018natural}. 
    The autoregressive architecture makes WaveNet capable of generating high-quality audio; however, it also leads to notably slow speed at inference time.  
      To address this problem, several methods are proposed.
  One is modifying the architecture of the autoregressive model and applying a more compact framework to reduce the computing time of generating each audio sample~\cite{jin2018fftnet, kalchbrenner2018efficient}. Without changing the nature of the autoregression, highly optimizing or weight pruning is still required to achieve real-time generation.

  Another approach to improve the autoregressive model is based on the teacher-student framework. 
  By applying knowledge distillation methods, a student model can learn from a well-trained teacher model to generate audio waveform in parallel. Although this framework has achieved remarkable real-time synthesis performance in \cite{oord2017parallel} and \cite{ping2018clarinet}, requirements of well-trained teacher models, highly-optimized distillation methods, and well-designed architectures remain problems for implementation.

  WaveGlow~\cite{prenger2019waveglow} is a non-autoregressive model that can generate high-quality audio samples in parallel. This flow-based neural vocoder is trained only to maximize the likelihood of the training data. The simple network and single cost function make it straightforward to implement and to train. The model is fast enough to real-time synthesize audio waveform. However, since WaveGlow is deep and contains a large number of parameters, it consumes huge computational resources during training and inference.

  In this paper, we aim to design an efficient, high-quality, and small footprint waveform generation model. We first apply the weight-sharing method to compress a WaveGlow and significantly reduce the size of the flow-based vocoder. A WaveNet-based post-filter is then applied to avoid the compression harming the speech quality. It is trained using loss functions on the frequency domains. Since the post-filter only needs to amend the output of the compressed WaveGlow, a small WaveNet is competent, keeping the overall model fast and lightweight. The proposed model, which we refer to as WG-WaveNet, possesses the advantage of simplicity in network architecture and loss function. Besides, compared with other neural vocoding methods, it requires much less computational cost during both training and inference.

  The contributions of this work are summarized as follow:
  \begin{itemize}
    \vspace{-2.9pt}
    \item  We propose a hybrid neural vocoder model, which is composed of a highly compressed WaveGlow model and a WaveNet-based post-filter. The proposed model, WG-WaveNet, is efficient and economical during training. WG-WaveNet can be trained on an NVIDIA 1080Ti GPU (using less than 8 GB GPU memory) in 4 days, while 8 NVIDIA GV100 GPUs were reported to use in the original WaveGlow paper~\cite{prenger2019waveglow}.
    \vspace{-2.9pt}
    \item  The proposed methods significantly improve the generating efficiency. In particular, the inference speed of the proposed WG-WaveNet is higher than 960 kHz using an NVIDIA 1080Ti GPU and 1.5 times faster than real-time even only using a CPU. 
    \vspace{-2.9pt}
    \item  For speech quality, perceptual experiments show that the proposed model can generate speech with a similar quality compared with WaveNet, WaveGlow, SqueezeWave~\cite{zhai2020squeezewave}, and Parallel WaveGAN~\cite{yamamoto2020parallel}.
    \vspace{-2.9pt}
    \item  We also study the quality of 44.1 kHz audio waveform (which we call high-fidelity audio) generated from neural vocoders. We explore the performances of recordings with various sampling rates and the effects of different parameters of short-time Fourier transform for training vocoders. The proposed method not only makes it possible to synthesize 44.1 kHz audio samples on a single CPU 1.2 times faster than real-time but also achieves a score of 4.01 in the MOS test, which even betters 16 kHz recordings.
  \end{itemize}
\vspace{-4pt}
\begin{figure*}[t]
  \centering
  \includegraphics[width=0.89\linewidth]{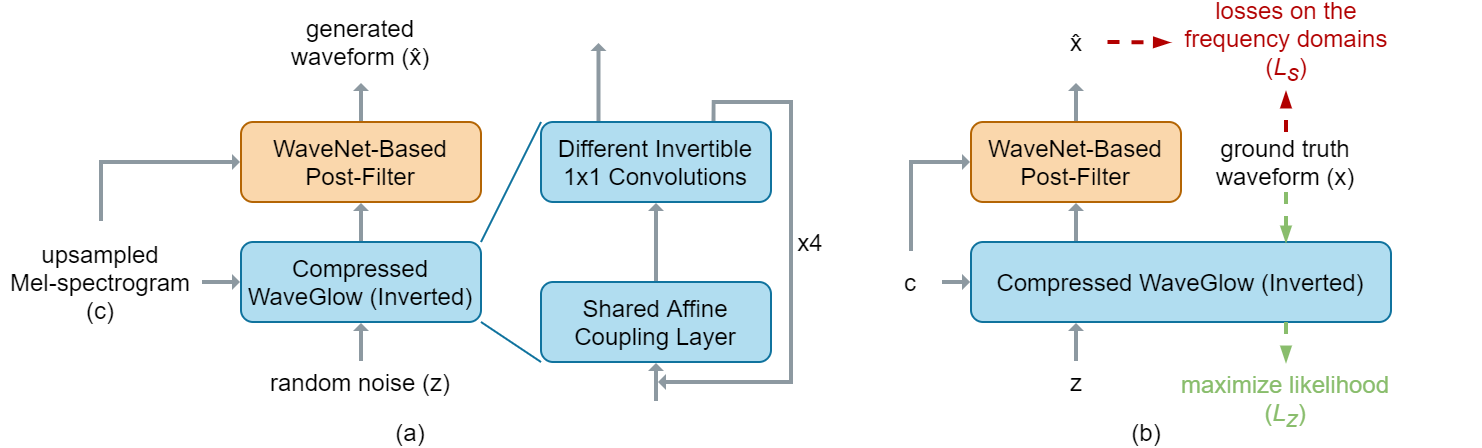}
  \vspace{-9pt}
  \caption{(a) The architecture of WG-WaveNet. (b) Training of WG-WaveNet.}
  \label{fig:model}
\vspace{-15pt}
\end{figure*}

\section{Proposed Model}

  The proposed WG-WaveNet is composed of two components, shown in Figure~\ref{fig:model}(a). The first part is a highly compressed WaveGlow model, which will be introduced in Section~\ref{sec:M1}. In Section~\ref{sec:M2}, to further improve the sound quality, we employ a WaveNet-based post-filter trained with loss functions on the frequency domains. 

\subsection{Highly Compressed WaveGlow}
\label{sec:M1}

  WaveGlow~\cite{prenger2019waveglow} is a reversible network trained for modeling the distribution of the real-world speech data. During training, the model takes audio samples as input and Mel-spectrograms as the condition. 
It learns to transform the distribution of the audio samples in the training dataset to a zero-mean spherical Gaussian. 
The training objective is to maximize the likelihood of the training data. During inference time, an inverted WaveGlow takes a randomly-sampled Gaussian noise as input and generates speech waveform conditioned on a Mel-spectrogram.

 The WaveGlow model consists of several transformations to progressively map speech data to Gaussian space. 
 A transformation is composed of an affine coupling layer~\cite{dinh2016density} and an invertible 1x1 convolution layer~\cite{kingma2018glow}. 
 Each affine coupling layer in WaveGlow adopts a WaveNet-like module. 
  The overall model is huge and hard to train.

We apply cross-layer parameter sharing to reduce parameters and make the model more compact. 
The cross-layer parameter sharing has shown to be helpful in NLP task pre-training~\cite{lan2019albert} and source separation~\cite{tuan2019mitas}.
As shown in Figure~\ref{fig:model}(a), transformations in the compressed WaveGlow share the same affine coupling layer
\footnote{To make the affine coupling layer shareable here,  
we remove the early-output mechanism used in the original WaveGlow to keep the output shape the same across layers.}. 
This approach keeps the model from drastically growing in size when it gets deeper. 
Considering that these transformations are processes of gradually mapping data from one distribution to another, invertible 1x1 convolution layers remain different across transformations to keep variability. We found that this improved the quality of generated speech in preliminary experiments. We also change the upsampling method from deconvolution to layers of duplication and convolution to further reduce parameters.

The training process is the same as mentioned in \cite{prenger2019waveglow}, shown by the green path in Figure~\ref{fig:model}(b). The loss function, denoted as $L_{z}$, is the negative log-likelihood of the training data. The proposed compression approach reduces the number of parameters in the WaveGlow and considerably cuts down the requirements of GPU memory. In the following section, we propose to use a post-filter to further speed up the convergence and improve the performance of the compressed WaveGlow.

\subsection{WaveNet-Based Post-Filter}
\label{sec:M2}


A random noise $z$ is sampled from a Gaussian as the input of the inverted compressed WaveGlow. The output of WaveGlow is then used as the input of the WaveNet-based post-filter to generates $\hat{x}$ in parallel~\cite{lluis2018end,rethage2018wavenet} conditioned on an upsampled Mel-spectrogram.
The WaveNet-based post-filter is trained by minimizing the loss function $L_{s}(x,\hat{x})$, in which $x$ is the ground truth samples, while $\hat{x}$ is the output of the post-filter.
The WaveNet-based post-filter and the inverted compressed WaveGlow are jointly learned to minimize $L_{s}(x,\hat{x})$\footnote{Since the WaveNet-based post-filter is irreversible, it can not be trained jointly by maximizing the likelihood as WaveGlow.}. 
Since the WaveNet here synthesizes audio samples based on the output of the inverted compressed WaveGlow, its parameters can also be highly reduced.
  
  For $L_{s}$, we utilize loss functions on the different frequency domains. 
  Spectral losses have been shown effective for training waveform generation models in \cite{takaki2019stft}, \cite{takaki2019training}, and \cite{yamamoto2020parallel}. We modify the multi-resolution Short-time Fourier transform (STFT) auxiliary loss in \cite{yamamoto2020parallel} as follows:
  \vspace{-6pt}
  \begin{equation}
    \label{sloss}
    L_{s}(x,\hat{x})=\frac{1}{M}\sum_{i=1}^{M}(L_{sc}^{i}(x,\hat{x})+L_{mag}^{i}(x,\hat{x})+L_{mel}^{i}(x,\hat{x})),
  \end{equation}
   where M is the number of different parameter sets of STFT; $L_{sc}$ and $L_{mag}$ are the spectral convergence loss and the log STFT-magnitude loss from \cite{arik2018fast}:
  \vspace{-6pt}
  \begin{equation}
    \label{scloss}
    L_{sc}(x,\hat{x})=\frac{\left \|\left |STFT(x)\right |-\left |STFT(\hat{x})\right |\right \|_{F}}{\left \|\left |STFT(x)\right |\right \|_{F}}
  \end{equation}
  \vspace{-6pt}
  \begin{equation}
    \label{magloss}
    L_{mag}(x,\hat{x})=\frac{1}{N_{mag}}\left \|log\left |STFT(x)\right |-log\left |STFT(\hat{x})\right |\right \|_{1},
  \end{equation}
  where $\left \|\cdot\right \|_{F}$ is the Frobenius norm, $\left \|\cdot\right \|_{1}$ is the $L_{1}$ norm, $\left |STFT(\cdot )\right |$ is the STFT magnitude, and $N_{mag}$ is the number of elements in the magnitude.
  To make $L_{s}$ more representative to human perception, we add a Mel-scale STFT-magnitude loss:
  \vspace{-6pt}
  \begin{equation}
    \label{melloss}
    L_{mel}(x,\hat{x})=\frac{1}{N_{mel}}\left \|log\left |MEL(x)\right |-log\left |MEL(\hat{x})\right |\right \|_{1},
  \end{equation}
  where $\left |MEL(\cdot )\right |$ and $N_{mel}$ denote the Mel-scaled STFT magnitude and the number of elements in the magnitude, respectively. The number of Mel bands differs in different STFT parameter sets.

  The WaveNet-based post-filter is trained jointly with the inverted compressed WaveGlow, as shown by the red path in Figure~\ref{fig:model}(b). The loss function for training WG-WaveNet is a linear combination of $L_{z}$ and $L_{s}$:
  \vspace{-6pt}
  \begin{equation}
    \label{all}
    L_{total}=\lambda L_{z}+L_{s},
  \vspace{-6pt}
  \end{equation}
  where $\lambda$ is a scalar to balance the loss terms. 
  In practice, $L_{s}$ is calculated every $n$ iterations.
  
  Eventually, the overall WG-WaveNet (compressed WaveGlow plus WaveNet postfilter) is \textit{one thirty-fifth} of the original WaveGlow in model size. Model details will be discussed in Section~\ref{sec:details}.

\section{Experiments}
\subsection{Datasets}
\label{sec:dataset}

  Two datasets were used in the experiments. One was the LJ Speech Dataset~\cite{ljspeech17}. This English dataset consists of 13100 clean audio clips (about 24 hours) of a female speaker. The sampling rate is 22050. The other was an internal Mandarin corpus, which contains 9004 utterances (about 6.8 hours) from a female speaker. The recordings were sampled at 44.1 kHz. 100 utterances were selected from each dataset for evaluation.
  
  We used the 80-band Mel-spectrogram as the condition to synthesize audio. For WG-WaveNet, the FFT size, hop size and window size for STFT are 2048, 200, and 800, respectively.

\subsection{Model Details}
\label{sec:details}

  The numbers of parameters of different models are listed in Table~\ref{tb:Model_Comparison}.
  The WaveNet-based post-filter in the proposed WG-WaveNet is composed of 7 layers of dilated convolution blocks with 64 channels. 
 The original WaveGlow has 12 transformations.
  With the help of the post-filter, the compressed WaveGlow consists of only 4 transformations. The WaveNet-like module in the shared affine coupling layer has 7 layers with 128 channels. 
  Both the WaveNet-based post-filter and the compressed WaveGlow have fewer numbers of layers and channels than the original WaveNet and WaveGlow, making WG-WaveNet much more compact. 
 WaveNet has 24.7 M parameters, and WaveGlow has 87.9 M parameters. 
 WG-WaveNet, on the other hand, has only 2.5 M parameters, which is 1/10 and 1/35 of those in WaveNet and WaveGlow, respectively.
  
  We compared our method with four different baseline models, WaveNet, WaveGlow, SqueezeWave, and Parallel WaveGAN. To ensure that the models were consistent compared to the original models, for the first three, we used pre-trained models from public implementations\footnote{https://github.com/r9y9/wavenet\_vocoder}\footnote{https://github.com/NVIDIA/waveglow}\footnote{https://github.com/tianrengao/SqueezeWave}. Note that the pre-trained models of WaveGlow and SqueezeWave were released by the official. We followed the setup in \cite{yamamoto2020parallel} to train the Parallel WaveGAN.
  The WG-WaveNet model was trained for 1 M steps using the Adam optimizer~\cite{kingma2014adam} with a batch size of 8. The learning rate was $4e^{-4}$ and reduced by half every 200 K steps. 
  We set $\lambda=1$ and $n=3$ based on preliminary experiments.
  The parameters for calculating $L_{s}$ in Section~\ref{sec:M2} are listed in Table~\ref{tb:fftp}.

  \begin{table}[h]
  \vspace{-6pt}
  \centering
  \caption{The parameters for calculating $L_{s}$.}
    \label{tb:fftp}
    \centering
  \vspace{-9pt}
  \begin{tabular}{|l|l|}
  \hline
  FFT size        & 4096, 2048, 1024, 512, 256 \\ \hline
  hop size        & 400, 200, 100, 50, 25      \\ \hline
  window size     & 1600, 800, 400, 200, 100   \\ \hline
  \# of Mel bands & 640, 320, 160, 80, 40      \\ \hline
  \end{tabular}
  \vspace{-15pt}
  \end{table}

\subsection{Speed and Computational Cost}
\label{sec:computing}
  We evaluated the speed and memory usage of different models during training and inference. Parallel WaveGAN and WG-WaveNet were trained on the same server using an Nvidia V100 16GB RAM GPU to fairly evaluate the computational cost at the training stage. The testing environment was a personal computer with an Intel i7-6700K CPU and an Nvidia 1080Ti GPU. Since the computational cost of parallel synthesis methods might be affected by the output length at the inference stage, we tested the models using utterances with various lengths uniformly distributed from 2 to 9 seconds.
  
  The results are shown in Table~\ref{tb:Model_Comparison}. 
  Though the training time of WG-WaveNet is slightly longer than that of Parallel WaveGAN, the training memory is 47\% less. 
  The inference speed of WG-WaveNet is at a rate of 967 kHz with GPU and 1.5 times faster than real-time without GPU.
  Moreover, we trained a faster version of WG-WaveNet, denoted as g-20. In WaveGlow, the input is reshaped to groups of 8 samples~\cite{prenger2019waveglow}. Inspired by \cite{zhai2020squeezewave}, the input of g-20 is reshaped to groups of 20 samples.
  This faster WG-WaveNet can be optimized with much less computational resources and generate 22 kHz speech 2.4 times faster than real-time without GPU. 

  \begin{table}[!t]
  \centering
  \caption{Comparison of computational cost and speed during training and inference time. The units of memory, time, and speed are GB, day, and kHz, respectively. Details of training and inference are described in Section~\ref{sec:computing}.}
    \label{tb:Model_Comparison}
    \centering
  \vspace{-9pt}
  \resizebox{80mm}{!}{
  \begin{tabular}{lccc}
  \toprule
  \textbf{Model} & \textbf{Size} & \textbf{\begin{tabular}[c]{@{}c@{}}Training\\ Mem. / Time\end{tabular}} & \textbf{\begin{tabular}[c]{@{}c@{}}Infer. Speed\\ CPU / GPU\end{tabular}} \rule{0pt}{2.5ex} \rule[-1.2ex]{0pt}{0pt} \\ \hline
  WaveNet           & 24.7 M & - & 0.1 / 0.12 \rule{0pt}{2.5ex} \\
  WaveGlow          & 87.9 M & - & 10 / 279 \rule{0pt}{2.5ex} \\
  SqueezeWave       & 23.7 M & - & 330 / 4486  \rule{0pt}{2.5ex} \\
  Parallel WaveGAN  & 1.3 M  & 14.4 / 2.7 & 18 / 841  \rule{0pt}{2.5ex} \\
  WG-WaveNet (ours) & 2.5 M  & 7.7 / 3.5 & 33 / 967 \rule{0pt}{2.5ex} \\
  WG-WaveNet (g-20) & 3.1 M  & \textbf{5.2} / \textbf{2.5} & \textbf{53} / \textbf{1634} \rule{0pt}{2.5ex} \rule[-1.2ex]{0pt}{0pt} \\ \bottomrule
  \end{tabular}
  }
  \vspace{-19pt}
  \end{table}

\subsection{Audio Quality Comparison}
\label{sec:quality}
  We conducted Mean Opinion Score (MOS) tests\footnote{Audio samples are publicly available at\\ https://bogihsu.github.io/WG-WaveNet/} as a subjective evaluation (higher is better) and calculated Mel Cepstral Distortion (MCD)~\cite{kubichek1993mel} as an objective evaluation (lower is better). In the MOS test, raters were asked to score utterances on a five-point scale according to the quality. Each utterance was randomly selected from the evaluation set and scored by at least 20 raters.
  
\begin{figure}[t]
  \centering
  \includegraphics[width=0.99\linewidth]{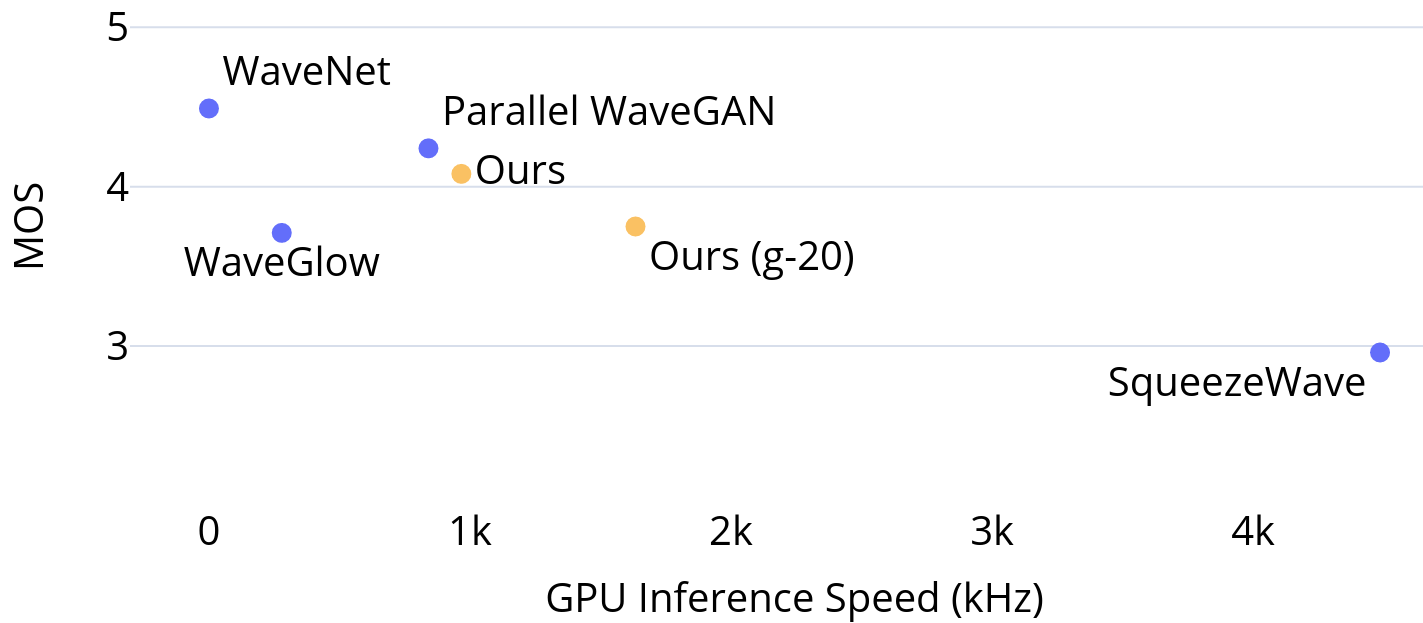}
  \vspace{-9pt}
  \caption{Trade-off between MOS and GPU inference speed.}
  \label{fig:tradoff}
\vspace{-19pt}
\end{figure}

  The evaluation results are shown in Table~\ref{tb:MOS_MCD}. To assess the effects of $L_{z}$ and $L_{s}$ on model performance, we trained WG-WaveNet with different $\lambda$ and $n$. 
  Figure~\ref{fig:tradoff} shows the trade-off between audio quality and inference speed.
  
  
  The observations based on Table~\ref{tb:MOS_MCD} and Figure~\ref{fig:tradoff} are concluded as follows: 
  (1) WaveNet has the highest MOS, which is close to that of the ground truth data, yet there is a gap between the performance of parallel and autoregressive synthesis methods. (2) MCD is not strongly related to human perception. Training a model using $L_{s}$ ($\lambda=0,n=1$) leads to the lowest MCD but not the best MOS, while WaveNet has the highest MOS and MCD. A similar contradictory result was also found in \cite{jin2018fftnet}. 
  (3) Though we used the official release models to synthesize utterances, WaveGlow and SqueezeWave did not perform well. Subjects reported there were noise and reverberation effects in the generated speech. 
  (4) The ablation study shows that both $L_{z}$ and $L_{s}$ are crucial for training WG-WaveNet. We found that training using only $L_{s}$ ($\lambda=0,n=1$) led to good quality at voiced part of speech but significant high-frequency glitch at unvoiced part. (5) The MOS decreases rapidly when the generating efficiency improves. WG-WaveNet, however, has the faster speed and a MOS of 4.08, which is close to that of Parallel WaveGAN. This indicates the proposed WG-WaveNet can greatly increase the synthesis speed while preserving a comparable performance.

  \begin{table}[t]
  \centering
  \caption{MOS and MCD results compared with other models. Mel-spectrograms were extracted from the ground truth. The MOS results are reported with 95\% confidence intervals.}
    \label{tb:MOS_MCD}
    \centering
  \vspace{-9pt}
  \begin{tabular}{lcc}
  \toprule
  \textbf{Model}                    & \textbf{MOS} & \textbf{MCD} \rule[-1.2ex]{0pt}{0pt} \\ \hline
  WaveNet                           & \textbf{4.49$\pm$0.101} & 4.619 \rule{0pt}{2.5ex} \\
  WaveGlow                          & 3.71$\pm$0.159 & 4.393 \rule{0pt}{2.5ex} \\
  SqueezeWave                       & 2.96$\pm$0.121 & 3.608 \rule{0pt}{2.5ex} \\
  Parallel WaveGAN                  & 4.24$\pm$0.108 & 4.026 \rule{0pt}{2.5ex} \\
  WG-WaveNet (ours)                 &            &       \rule{0pt}{2.5ex} \\
  \hspace*{0.3cm} $\lambda=1,n=3$   & 4.08$\pm$0.118 & 3.783 \rule{0pt}{2.5ex} \\
  \hspace*{0.3cm} $\lambda=1,n=1$   & 3.23$\pm$0.159 & 2.948 \rule{0pt}{2.5ex} \\
  \hspace*{0.3cm} $\lambda=0,n=1$   & 3.65$\pm$0.164 & \textbf{2.407} \rule{0pt}{2.5ex} \\
  \hspace*{0.3cm} g-20              & 3.75$\pm$0.124 & 3.848 \rule{0pt}{2.5ex} \\ \hline
  Ground Truth                      & 4.61$\pm$0.096 & - \rule{0pt}{2.5ex} \\ \bottomrule
  \end{tabular}
  \vspace{-10pt}
  \end{table}

\subsection{High-Fidelity Audio Generation}

  Due to the fast inference speed of WG-WaveNet as shown in Sections~\ref{sec:computing} and~\ref{sec:quality}, we show that WG-WaveNet can generate high-fidelity audio (44.1 kHz) in this subsection.
  To evaluate the performance, we trained WG-WaveNet and Parallel WaveGAN on the 44.1 kHz speech dataset mentioned in Section~\ref{sec:dataset}. 
   We only compared WG-WaveNet with Parallel WaveGAN here because only Parallel WaveGAN and SqueezeWave are efficient enough to synthesize 44.1 kHz audio, and the audio quality of SqueezeWave is not comparable with Parallel WaveGAN. 
  MOS tests with the same setups as those in Section~\ref{sec:quality} were conducted on the generated waveform and ground truth data with different sampling rates.
  
  The results are shown in Table~\ref{tb:44KC}. 
  "w800" denotes that the window size for extracting Mel-spectrograms is set to 800. 
  The FFT size, hop size, and the number of Mel bands are also the same as mentioned in~\ref{sec:dataset}. 
  "w1600" denotes that the window size is doubled to 1600, and the other parameters are also doubled. 
  Since the sampling rate is changed from 22050 to 44100, doubling STFT parameters (w1600) makes the temporal resolution of extracted features the same as in Section~\ref{sec:quality}, while the temporal resolution is doubled in "w800".
  Similarly, parameters for calculating $L_{s}$ in "w800" are the same as in Table~\ref{tb:fftp}, while they are doubled in "w1600".
  We first found that the sampling rates of the ground truth samples significantly affect their perceptual scores. 
  The raters considered the ground truths with higher sampling rates are better.
  Experiments reveal that when the temporal resolution of acoustic features is fixed (w1600), it is harder to generate 44.1 kHz speech than to generate the 22 kHz one.
  We observed that Mel-spectrograms with higher temporal resolution (w800) helped improve the performance of WG-WaveNet (w800). 
  WG-WaveNet outperformed Parallel WaveGAN in both "w800" and "w1600" cases.
  Eventually, the faster WG-WaveNet reached 4.01 MOS, which is even better than that of 16 kHz ground truth speech.

\subsection{Text-to-Speech}

  We combined WG-WaveNet with a Tacotron 2 model to evaluate the proposed method as a vocoder. The Tacotron 2 was built following \cite{shen2018natural}. Data preprocessing for training the TTS model and vocoders were set to the same as mentioned in Section~\ref{sec:dataset}.
  
  The results of MOS tests and GPU inference speed of vocoders are reported in Table~\ref{tb:TTS}. Note that the ground truth inherently has better prosody and quality than those of the speech generated by Tacotron 2. We found that the performance gap between WaveNet and the parallel synthesis methods narrowed. WG-WaveNet has the MOS comparable to that of Parallel WaveGAN, and the inference speed is faster than other methods, which shows the advantage of WG-WaveNet as a vocoder for fast high-quality speech synthesis. 

  \begin{table}[t]
  \centering
  \caption{MOS results of high-fidelity audio generation with 95\% confidence intervals. Mel-spectrograms were extracted from the ground truth sampled at 44.1 kHz.}
    \label{tb:44KC}
    \centering
  \vspace{-9pt}
  \begin{tabular}{lc}
  \toprule
  \textbf{Model}                    & \textbf{MOS} \rule[-1.2ex]{0pt}{0pt} \\ \hline
  Parallel WaveGAN                  &            \rule{0pt}{2.5ex} \\
  \hspace*{0.3cm} w1600         & 3.12$\pm$0.134 \rule{0pt}{2.5ex} \\
  \hspace*{0.3cm} w800          & 3.04$\pm$0.126 \rule{0pt}{2.5ex} \\
  WG-WaveNet (ours)                 &            \rule{0pt}{2.5ex} \\
  \hspace*{0.3cm} w1600          & 3.15$\pm$0.148 \rule{0pt}{2.5ex} \\
  \hspace*{0.3cm} w800          & 3.71$\pm$0.131 \rule{0pt}{2.5ex} \\
  \hspace*{0.3cm} w800 (g-20)   & \textbf{4.01$\pm$0.110} \rule{0pt}{2.5ex} \rule[-1.2ex]{0pt}{0pt} \\ \hline
  Ground Truth (16 kHz)             & 3.72$\pm$0.147 \rule{0pt}{2.5ex} \\
  Ground Truth (22 kHz)             & 4.15$\pm$0.127 \rule{0pt}{2.5ex} \\
  Ground Truth (44.1 kHz)           & 4.44$\pm$0.105 \rule{0pt}{2.5ex} \\ \bottomrule
  \end{tabular}
  \vspace{-8pt}
  \end{table}
  
  \begin{table}[t]
  \centering
  \caption{MOS results and GPU inference speed (in kHz) compared with other models. Mel-spectrograms were generated by the Tacotron 2 model. The MOS results are reported with 95\% confidence intervals.}
    \label{tb:TTS}
    \centering
  \vspace{-9pt}
  \begin{tabular}{lcc}
  \toprule
  \textbf{Model}                & \textbf{MOS}  & \textbf{\begin{tabular}[c]{@{}c@{}}Infer.\\ Speed\end{tabular}} \rule[-1.2ex]{0pt}{0pt} \\ \hline
  Tacotron 2+GL                 & 2.11$\pm$0.139    & - \rule{0pt}{2.5ex} \\
  Tacotron 2+WaveNet            & \textbf{3.96$\pm$0.116} & 0.12 \rule{0pt}{2.5ex} \\
  Tacotron 2+Parallel WaveGAN   & 3.72$\pm$0.127    & 841 \rule{0pt}{2.5ex} \\
  Tacotron 2+WG-WaveNet (ours)  & 3.68$\pm$0.133    & \textbf{967} \rule{0pt}{2.5ex} \rule[-1.2ex]{0pt}{0pt} \\ \hline
  Ground Truth                  & 4.36$\pm$0.108    & - \rule{0pt}{2.5ex} \\ \bottomrule
  \end{tabular}
  \vspace{-19pt}
  \end{table}

\section{Conclusion}

  We proposed WG-WaveNet, a fast, lightweight, and high-quality waveform generation model.
  Combining with a highly compressed WaveGlow and a WaveNet-based post-filter, WG-WaveNet requires much less computational resources compared to other parallel synthesis methods during both training and inference time.
  The experimental results show that WG-WaveNet is capable of generating high-quality 22 kHz and 44.1 kHz audio samples faster than real time without GPU.


\bibliographystyle{IEEEtran}
\bibliography{mybib}
\end{document}